\documentclass[aps,twocolumn,showpacs]{revtex4}
\usepackage{graphicx}
\usepackage{textcomp}
\usepackage{psfrag}

\begin{document}
\title{Dynamics of the peel front and the nature of acoustic emission during peeling of an adhesive tape}
\author{Rumi De$^{1,\star}$}
\author{G. Ananthakrishna$^{1,2,\dag}$}
\affiliation{$^1$ Materials Research Centre, Indian Institute of Science,
Bangalore-560012, India.\\
$^2$ Centre for Condensed Matter Theory, Indian Institute  of Science,
 Bangalore-560012, India.}
 
\begin{abstract}
We investigate the peel front dynamics and acoustic emission of an 
adhesive tape within the context of a recent model by including an 
additional dissipative energy that mimics bursts of acoustic signals.  We find 
that the nature of the peeling front can vary from smooth to stuck-peeled 
configuration depending on the values of dissipation coefficient, inertia of 
the roller, mass of the tape.  Interestingly, we find that the distribution of AE 
bursts shows a power law statistics with two scaling regimes with increasing pull velocity  as observed in experiments. In these regimes, the stuck-peeled configuration is similar to the `edge of peeling' reminiscent of a system driven to a critical state.
\end{abstract}
 

\maketitle
The process of peeling of an adhesive material from a substrate is a 
complicated phenomenon involving molecular attraction at the interface, and 
 kinetic and dynamical effects. The kinetic nature of the process is clear 
from the fact that the peel force depends on the peel rate. The 
fracture process during peeling  can be either cohesive or adhesive at low and high peel velocities 
respectively. At intermediate velocities,  the 
fracture process is intermittent suggesting that the peeling process 
 results from an interplay of time scales. At low peel velocities, 
there is sufficient time for  viscoelastic glue to fully relax while at high 
velocities, the glue essentially behaves like a solid \cite{deGennes96}. The intermittent behavior is observed 
when  the viscoelastic time scale is of the same order as the peel rate time scale. It is an everyday experience that  the peeling  process is always  accompanied by  a characteristic  audible noise \cite{MB,Ciccotti04}.  
However, the mechanism leading to the acoustic emission (AE) has remained ill 
understood. Moreover, the inhomogeneous deformation of 
the peel front results from the destabilization of uniformly advancing peel 
front \cite{GL}. To the best of our knowledge, we are not aware of any model 
that investigates the dynamics of the peeling front and the associated 
acoustic emission.  We address these two issues within the context of a 
model for the peeling  of an adhesive tape.

Experiments on peeling of an adhesive tape mounted on a  roller \cite{MB} show that the peel force function has two stable branches separated  by an unstable one.  The pull force exhibits a rich variety of behavior  ranging from sawtooth to irregular   waveforms  \cite{MB,HY,Ciccotti98}.  A dynamical analysis of the force waveforms and the AE  signals reports chaotic dynamics at the upper end of  pull velocities  \cite{Gandur}. However, as there are no models, no further insight into the origin of acoustic emission has been possible.

A  relevant model introduced in \cite{MB} has been
studied by  others (Ref. \cite{HY,Ciccotti98,Ciccotti04}) belongs to 
the category of differential algebraic equations (DAE) and are singular 
requiring an appropriate DAE algorithm provided in Ref. \cite{Rumi04}. 
[For this reason, the results in Ref. \cite{HY} are the artifact of the 
method followed.] Recently, we modified these equations  into a set of 
ordinary differential equations (ODE) by including  the kinetic energy of 
the stretched tape. The ODE model not only supports  dynamical jumps across 
the two stable branches, it displays a rich  dynamics \cite{Rumi05}. Here, 
we extend this model to include spatial degrees of freedom to study the 
contact line dynamics of the peeling front. The inclusion of a local strain 
rate dependent Rayleigh dissipation functional along with 
the kinetic  energy of the tape forms a basis for converting the 
potential energy stored in the stretched tape into kinetic energy provides a mechanism for explaining  qualitative experimental features on acoustic 
emission (AE) \cite{Rumi04a, Rajeev}.

\begin{figure}[!t]
\hbox{
\includegraphics[height=2.0cm, width=5.0cm]{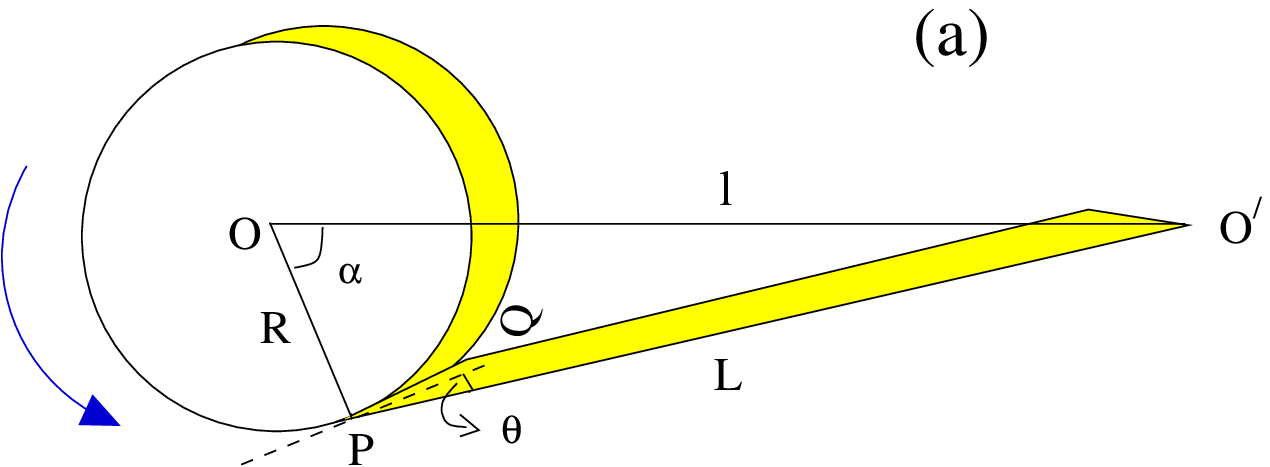}
\psfrag{vs}{{$v^s$}}
\psfrag{phixxxx}{$\phi(v^s)$}
\includegraphics[height=2.5cm,width=3cm]{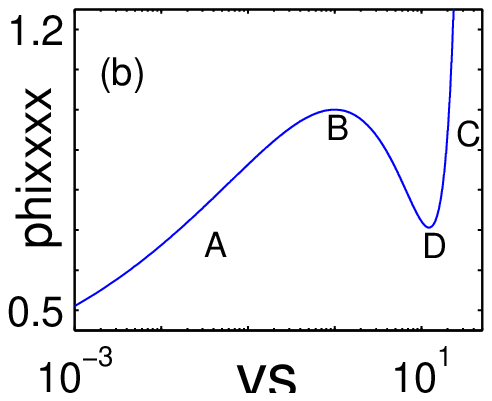}}
\caption{(a) A schematic plot of the experimental setup. (b) Plots of $\phi(v^s)$ as a function of $v^s$ for $V^s=1.45$.}
\label{tapewidth}
\end{figure}

Figure \ref{tapewidth}(a)  shows a schematic representation of the experimental  setup. An adhesive roller tape of radius $R$ is mounted on an axis passing through $O$  and is pulled at a constant speed $V$ by 
a couple meter motor positioned at $O'$. Then, the line  $PQ$  represents 
the peeling front. Several features of  the setup can be explained by considering the 
projection on to the plane of the paper ($OPO'$). The tangent to the contact 
point $P$  (representing  the contact line $PQ$)  subtends an angle 
$\theta$ to the line  $PO^{\prime}$. Let the distance between  $O$ to  
$O'$ be $l$ and the peeled length of the tape $PO^{\prime}$ be $L$. If $P$ 
subtends an angle $\alpha$ at $O$ with the horizontal  $OO'$, the geometry 
of the setup gives $L\ {cos}\, \theta = -l\ {sin}\,\alpha$ and 
$L\ {sin}\,\theta = l\ { cos}\,\alpha - R$. As the local velocity $v$ at
$P$  undergoes rapid bursts during rupture,  we 
have  $V= v + {\dot u} - \dot L= v + \dot u + R \ \ {\rm cos}\ \theta  \ \dot \alpha$. Let  $u(y)$ to be displacement with respect to the uniform  `stuck' peel front and let  $v(y),\theta(y),\alpha(y)$ be defined at every point $y$ along the contact line. As  the tape of width $b$ is pulled with a velocity $V$, the above equation generalizes to 
\begin{eqnarray}
{1\over b} \int^b_0 \big[V- v(y) -  \dot u(y)  -  R \ \ {\rm cos} \ \ \theta(y) \ \ \dot\alpha(y)\big]dy =0. 
\label{Vconstraint}
\end{eqnarray}
However, as we are interested in the dynamics of the contact line of the softer glue material (whose elastic constant is three orders less than that of tape material), the effective spring constant $k_{g}$ of the contact line is assumed to be much less than that of the tape  material $k_t$. Then, as the entire tape is pulled with a velocity $V$, the  force along $PO^{\prime}$ equilibrates  fast, we can assume that the integrand in Eq. (\ref{Vconstraint})  is zero for all $y$.

We derive the equations of motion  of the contact line by considering the 
Lagrangian ${\cal L}= U_K - U_P$, where $U_K$  and $U_P$ are  the kinetic and potential energies respectively. The kinetic energy is given by
$U_K={1\over2}\int^b_0 \xi \big[\dot \alpha(y) +{v(y)\over R} \big]^2 dy + {1\over2}\int^b_0 \rho \big[\dot u(y) \big]^2 dy$, where
the first term represents the rotational kinetic energy of the roller tape
and second term arises due to the kinetic energy of the stretched part of the 
tape. Here, $\xi$ is the moment of inertia per unit width of the roller tape
and $\rho$  the mass per unit width of the length $L$. The total potential energy (PE) $U_P$ of the stretched ribbon can be written as $U_P={1\over2}\int^b_0 {k_t\over b} \Big[u(y) \Big]^2 dy + {1\over2}\int^b_0 {k_g b} \Big[{\partial u(y) \over \partial y} \Big]^2 dy$. 
The total dissipation has two contributions 
${\cal R}={1\over b} \int^b_0 \int f(v(y)) dv dy + {1\over2}\int^b_0 {\Gamma_u\over b} \Big[{\partial \dot u(y) \over \partial y} \Big]^2 dy$,
where $f(v,V)$ physically represents the peel force function assumed to be derivable from a potential function  $\Phi(v) = \int f(v)dv$ (see Ref. \cite{Rumi05}). The second term,  denoted by $ R_{ae}$, represents the dissipation 
arising from the rapid movement of peel front is given by  the Rayleigh 
dissipative  functional. This has the same form as the energy 
dissipated in the form of acoustic emission during abrupt motion of 
dislocations in plastic deformation, 
i.e.,  $E_{ae} \propto \dot \epsilon^2(r)$, where $\dot \epsilon(r)$ is 
the local plastic strain rate \cite{Rumi04a}. Hence we interpret 
$R_{ae}$ as the energy dissipated in the form of AE signals. Indeed, 
such a term has been successfully used to explain several features of AE 
signals in martensites \cite{Rajeev}. 

We  rewrite all the energy terms in a scaled form using  a time like variable $\tau = \omega_{u} t$ where $\omega_{u}^2={k_t/(b \ \rho)}$ . Let $f_{max}$ and $v_{max}$ the maximum value of  $f(v)$ and $v$ on the left 
stable branch.  Then,  defining a length scale  $d=f_{max}/k_t$,  we introduce $u = X d = X (f_{max}/k_t) $, $l =  l^s d$, $L =  L^s d$   and $R =  R^s d$. The peel force $f$ can be written as 
$f=f_{max}\phi(v^s)$  where $v^s=v/v_c\omega_u d$  
and $V^s=V/v_c\omega_u d$ are the dimensionless peel and pull velocities 
respectively.  Here, $v_c = v_{max}/ \omega_u d$ is dimensionless critical 
velocity at which the unstable branch starts (in the scaled units, 
the unstable branch begins at $v^s =1$). Defining  
$C_f=(f_{max}/k_t)^2(\rho/\xi)$,   $k_0=k_g b^2/(k_t a^2)$,  
$\gamma_u = \Gamma_u \omega_u/(k_t a^2)$, and $y = ar$, where $a$ is a 
unit length variable along the peel front, the scaled local form of 
Eq. (\ref{Vconstraint}) is 
\begin{equation}
\dot X = (V^s - v^s)v_c + R^s \ {l^s \over L^s} \ ({sin}\ \alpha)\ \dot \alpha 
\label{localconstraint}
\end{equation}  
The scaled kinetic and potential energies can be written as $U^s_K = {1\over2 C_f}\int^{b/a}_0 \Big[\dot \alpha(r) +{v_c v^s(r)\over R^s} \Big]^2 dr + {1\over2}\int^{b/a}_0 \Big[\dot X(r) \Big]^2 dr$ and 
$U^s_P = {1\over2}\int^{b/a}_0 X^2(r)  dr + {k_0\over2}\int^{b/a}_0 \Big[{\partial X(r) \over \partial r} \Big]^2 dr $ 
respectively. The total dissipation in a scaled form is
${\cal R}^s = {1\over b} \int^{b/a}_0 \int \phi(v^s(r)) dv^s dr + {1\over2}\int^{b/a}_0 \gamma_u \Big[{\partial \dot X(r) \over \partial r} \Big]^2 dr$.
$\phi(v^s)$ is the scaled peel force that can be obtained by using in Eq. (9) of Ref. \cite{Rumi05} shown in Fig.  \ref{tapewidth}(b). We shall refer the left branch AB as the `stuck state' and  the high velocity branch CD as the peeled state.

Using ($\alpha(r), \dot\alpha(r), X(r),\dot X(r)$) as generalized 
coordinates in the Lagrange equations of motion, 
${d\over d\tau}\left({\partial{\cal L} \over {\partial\dot\alpha(r)}}\right)-
{\partial{\cal L} \over {\partial\alpha(r)}}+
{\partial{\cal R}^s \over {\partial\dot\alpha(r)}}=0$,
and
${d\over d\tau}\left({\partial{\cal L} \over {\partial \dot X(r)}}\right)-
{\partial{\cal L} \over {\partial X}(r)}+
{\partial{\cal R}^s \over {\partial \dot X(r)}}=0$,
we get the equations of motion as
\begin{eqnarray}
 \ddot \alpha &=& - {v_c \dot v^s \over R^s}  - C_f R^s {l^s/L^s \, {sin}\, \alpha \over (1 + l^s/L^s \, {sin}\, \alpha)} \phi(v^s),\label{seqalpha} \\
\ddot X &=& - X + k_0 {\partial^2 X \over \partial r^2}  + {\phi(v^s)\over (1+ l^s/L^s \, {sin}\, \alpha)} + \gamma_u {\partial^2 \dot X \over \partial r^2}  
\label{sequ}
\end{eqnarray}
Equations (\ref{seqalpha},\ref{sequ}) are still not suitable for further analysis as  they have to satisfy  the constraint equation Eq.(\ref{localconstraint}). In the spirit of  mechanical  systems with constraints \cite{ECG}, we obtain the equation for the acceleration  variable $\dot v^s (r)$  by differentiating Eq. (\ref{localconstraint}) to be
\begin{eqnarray}
 \dot v^s = \big[- {\ddot  X} + {R^sl^s\over L^s} \big(\dot \alpha^2 (cos \alpha -{R^s l^s}({sin \alpha \over L^s})^2  \big) 
 +  sin \alpha {\ddot \alpha} \big) \big]/v_c.
\label{sdotv}
\end{eqnarray}

Eqs. (\ref{localconstraint},\ref{seqalpha}) and (\ref{sdotv})  were solved by discretizing and using an adaptive step size stiff differential equations solver (MATLAB package) for  open boundary conditions. The initial conditions were chosen from the stuck state [ ie., AB branch of $\phi(v^s)$] with a small spatial inhomogeneity in $X$ that satisfies Eq.(\ref{localconstraint}) approximately. The system is evolved till a steady state is reached before the data is accumulated. We have studied the dynamics  over a wide range of  values of $C_f$, $V^s$, and $\gamma_u$ keeping $R^s=0.35$, $l^s=3.5 $, $k_0=0.1$, $N=50$  and $N =100$. Note  that $v_c$ is an important parameter which however is determined once $f(v)$ is given. (The values of the unscaled parameters, for example $k_t \sim 1000$,   are fixed using the data in Ref. \cite{Ciccoae}.   $f(v)$ used here preserves major the features of the  experimental curve such as $f_{max} \sim 280 N/m$ at $v_{max} = 0.05 cm/s$ with a velocity jump to 16 cm/s. See also \cite{Rumi04}. Note however, we do not use the dynamization scheme used in \cite{Rumi04}.)  These equations exhibit  rich dynamics which can be classified as uniform,   rugged and stuck-peeled nature of the contact line.  In the unscaled variables, the results reported here correspond to changing the tape mass ($m = \rho b$) keeping $I (=\xi b)$ constant.
\begin{figure}[]
\hbox{
\includegraphics[height=3.5cm,width=4cm]{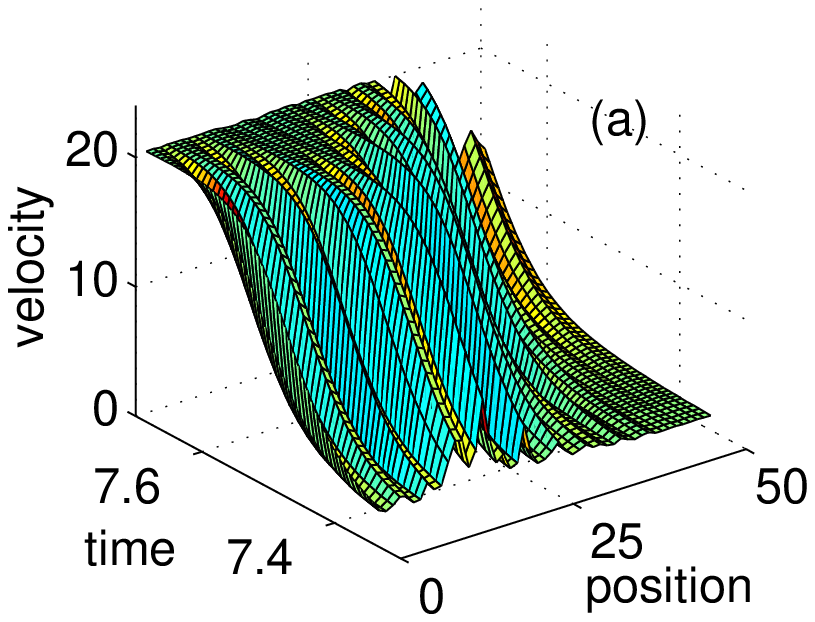}
\includegraphics[height=3.5cm,width=4cm]{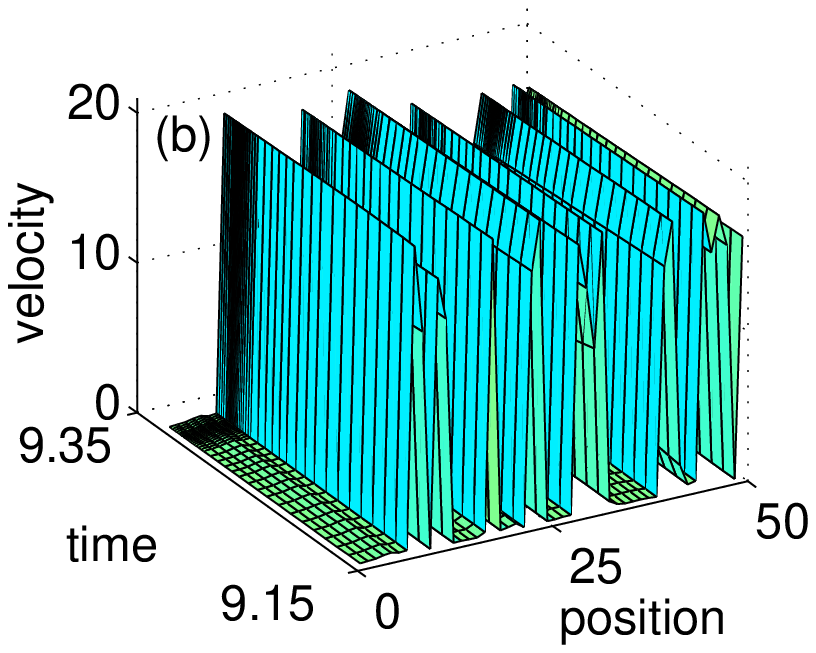}
}
\caption{Plot of the peel velocity configuration  for $C_f = 0.00788$ 
and $V^s=1.45$. 
(a) A rugged peel front for $\gamma_u=0.1$.
(b) Illustrates a stuck-peeled configuration  for $\gamma_u=0.01$.
}
\label{I2m3V1gu1p1}
\end{figure}

First consider the results for $C_f = 0.00788$ $(v_c= 0.0024) $ and $V^s=1.45$  (i.e., high inertia, low mass, and low pull velocity regime in the unscaled parameter space) as we decrease the dissipation parameter $\gamma_u$ from 1.0 to 0.001.   We observe a wide variety of events, some of which are illustrated in the  plots of the peel velocities $v^{s}_{i}(\tau)$. For instance, all spatial points peel together  for  $\gamma_{u}=1.0$.  As we decrease $\gamma_{u}$ to $0.1$ keeping other parameters fixed, the contact line profile becomes rugged even though all  points  peel nearly at the same time as seen in Fig. \ref{I2m3V1gu1p1}(a). 

Intuitively, high $\gamma_u$ implies that velocities of neighboring points  are coupled strongly and hence are not allowed to follow their local site dynamics. Thus, the total dissipation ${R}_{ae}^s(\tau)={1\over 2}\gamma_u \sum_i(\dot X_{i+1}- \dot X_i)^2$ is vanishingly small when peeling is coherent. In contrast, for low $\gamma_u$ [ say 0.1 as in Fig. \ref{I2m3V1gu1p1}(a)], the coupling between neighboring velocities is weak and the local dynamics dominates. This means  more ruggedness and hence higher dissipation.

\begin{figure}[!t]
\hbox{
\includegraphics[height=2.8cm,width=3cm]{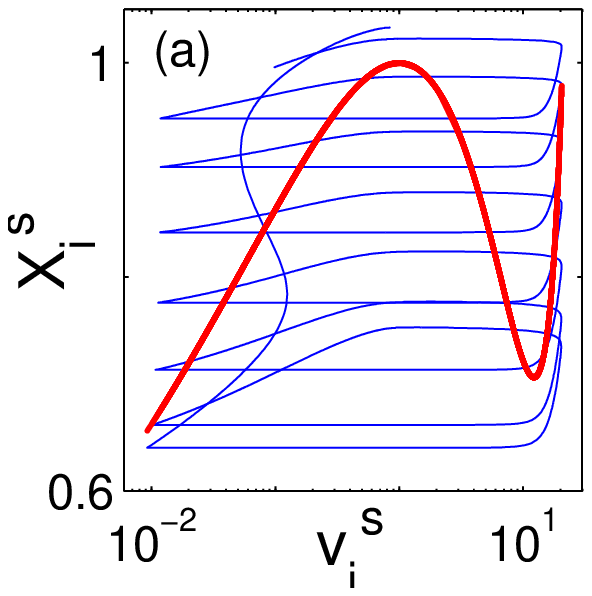}
\includegraphics[height=3.2cm,width=5cm]{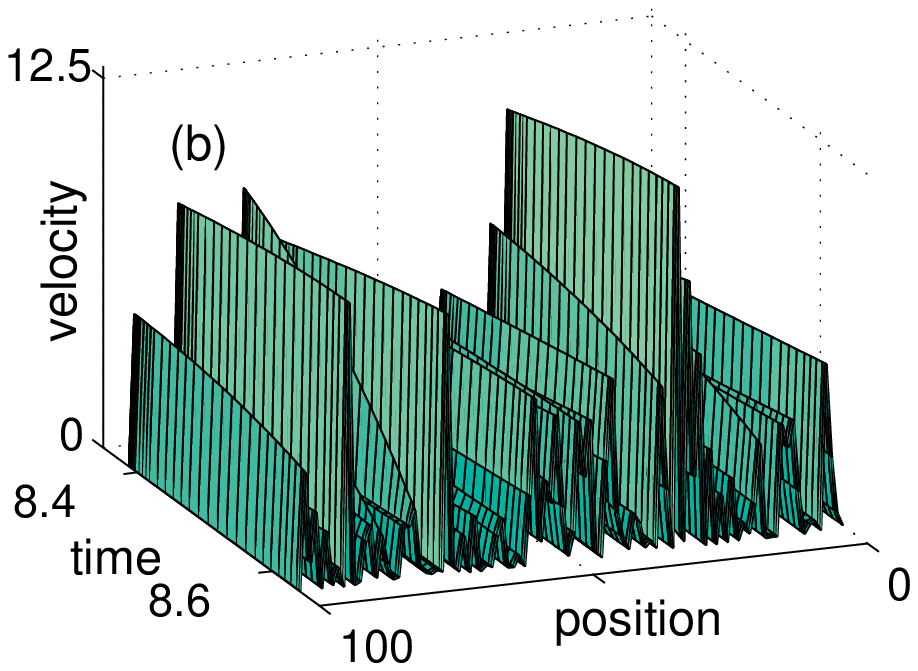}}
\vspace*{0.1cm}
\hbox{
\includegraphics[height=2.8cm,width=5cm]{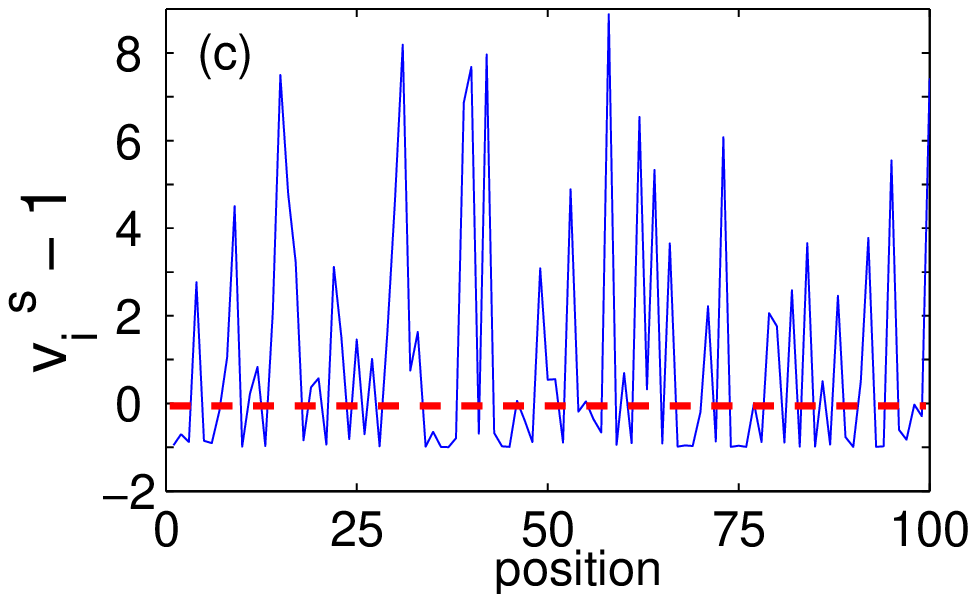}
\includegraphics[height=2.8cm,width=3cm]{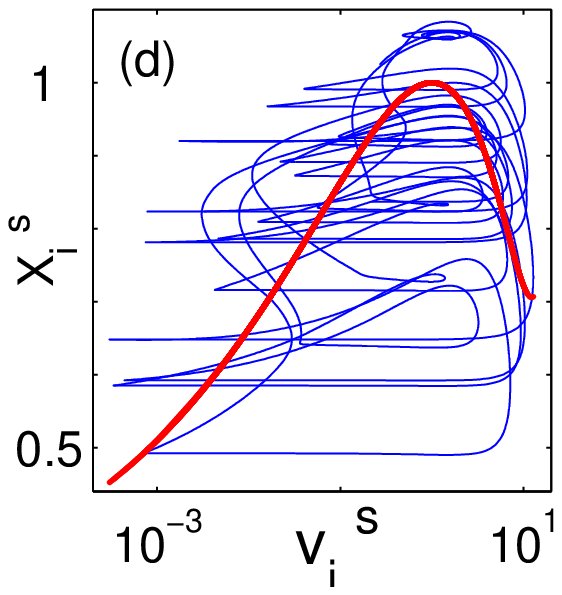}}
\caption{
(a) The phase plot for  $i =25$ for $C_f = 0.00788$ and $V^s=1.45$, and $\gamma_u=0.01$. 
(b) Plot of `an edge of peeling' configuration for $C_f = 0.788$ and $V^s=1.45$, and $\gamma_u=0.01$.  
(c) The corresponding instantaneous plot of `an edge of peeling' configuration. Dashed line represents the critical peel 
velocity $v_{max}^s = 1$. (d) The corresponding  phase plot for $i =25$. Bold lines in (a) and (d)  represent $\phi(v_{i}^s)$.
}
\label{I2m1V1gup01}
\end{figure}

As $\gamma_{u}$ is decreased to 0.01, the peel front exhibits two types of configurations depending on whether the system is on AB branch entirely,  or partly on CD and AB branches of $\phi$. When on CD branch,  the ruggedness is substantially higher than that for $\gamma_u = 0.1$ (Fig. \ref{I2m3V1gu1p1}(a)). Once the peeling process starts, the peel front breaks up into regions of stuck and peeled segments as  shown in  Fig. \ref{I2m3V1gu1p1}(b).   This configuration results from the orbit jumping between the low and high velocity branches of  $\phi(v^s)$. A  typical phase plot of  $X_i^s$ versus $v_i^s$  is shown in Fig. \ref{I2m1V1gup01}(a) for $i=25$ with other points differing only in phase. Thus, as the phase difference along the peel front builds up to a value equal to the phase difference between the orbits that are in the stuck and the peeled states, the stuck state changes to a peeled state or vice versa.

\begin{figure}[!t]
\hbox{
\includegraphics[height=2.9cm,width=4cm]{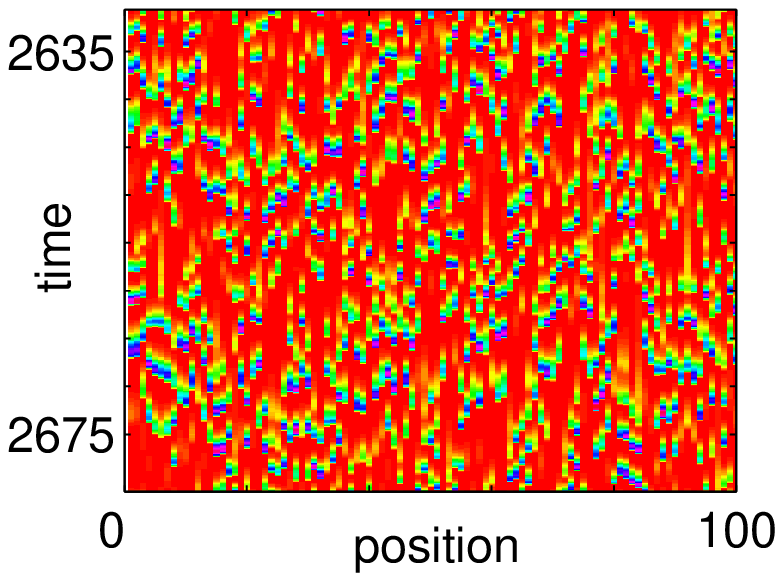}
\psfrag{Rsae}{{$R^s_{ae}$}}
\includegraphics[height=2.9cm,width=4cm]{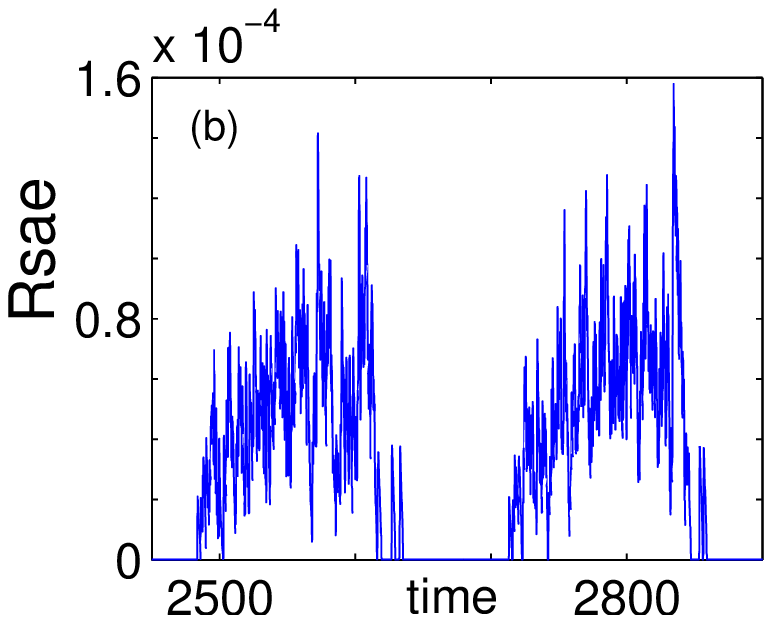}
}
\vspace*{0.1cm}
\hbox{
\psfrag{Rsae}{{$R^s_{ae}$}}
\includegraphics[height=2.9cm,width=4.0cm]{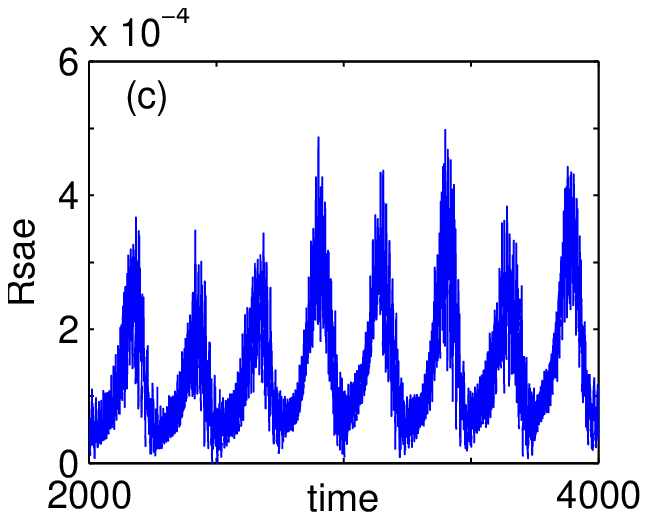}
\includegraphics[height=2.8cm,width=4.cm]{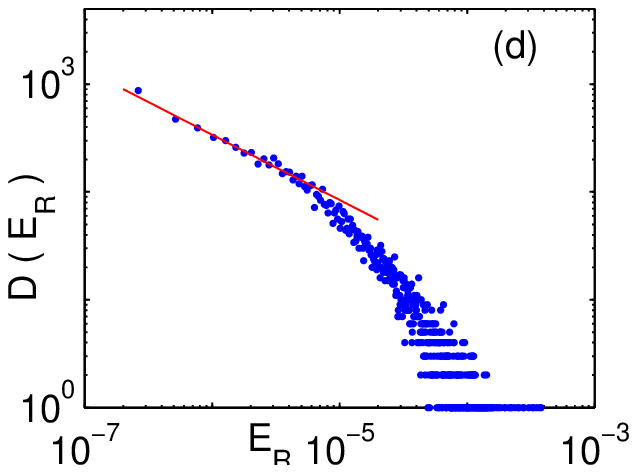}
}
\caption{
(a) Spatio-temporal chaotic plot for $C_f = 0.788, V^s = 1.45$, 
and $\gamma_u =0.01$ (color on line, low  $\rightarrow$  high peel velocity: 
red $\rightarrow$ yellow $\rightarrow$ green $\rightarrow$ blue $\rightarrow$ 
pink). (b) Plots of $R_{ae}^s(\tau)$ {\it vs.} time $\tau$ for $C_f =0.00788,V^s =1.45,\gamma_u= 0.01$.
(c) $R_{ae}^s(\tau)$ for $C_f = 0.00788,V^s = 5.93$, 
$\gamma_u =0.01$ and (d) the corresponding distribution $D(E_R)$ of the 
amplitudes $E_R$  showing two scaling regimes. 
}
\label{4AEI2m3m1V1gup1}
\end{figure}

Now consider the dynamics for higher mass ($C_f = 0.788, v_c =0.024$) 
and $V^s=1.45$ as we  decrease $\gamma_u=1.0$ to $0.001$.  For this entire range of $\gamma_u$, the peel front displays stuck-peeled segments for all times as shown in Fig. \ref{I2m1V1gup01}(b). There is a dynamic equilibrium between the peeled and stuck segments with the segments that are  stuck at some instant getting  unstuck at another instant and vice versa. Further, as is clear  from Fig. \ref{I2m1V1gup01}(b),  the average  of the velocity jumps along the peel front is smaller than the low mass case (compare Fig. \ref{I2m3V1gu1p1}).  Concomitantly, the number of stuck segments increases with each stuck segment having a only few stuck points better illustrated in an instantaneous  plot of $v^s_i-1$ shown in Fig. \ref{I2m1V1gup01}(c).   Moreover, from Fig. \ref{I2m1V1gup01}(c), it is clear that even the points that are in the stuck state are barely stuck. Further,  it is clear that  the orbit spends considerable time around the maximum which is the critical peel value [Fig. \ref{I2m1V1gup01}(d)].  Thus,  Figs.\ref{I2m1V1gup01}(b) and \ref{I2m1V1gup01}(c) correspond to a verge of peeling state.  The 'edge of peeling' picture  remains unaltered with time even though the stuck points themselves change.    The largest Lyapunov exponent is 0.15 ( for $N=50$) and hence this state is spatio-temporally chaotic [Fig. \ref{4AEI2m3m1V1gup1} (a)].

In experiments, the nature of the AE signals changes from  burst type to continuous type as the pull velocity is increased. In the model, the rate of dissipated energy $R_{ae}^s = -dE_{ae}/d\tau $ represents the AE bursts.  We have studied  the statistics of ${R}_{ae}^s$ as we increase the pull velocity keeping the tape mass low ($C_f= 0.00788,v_c=0.0024$).    As in experiments, for small $\gamma_u$, at low  velocities,  we find that  ${R}_{ae}^s$  exhibits bursts followed by a quiescent state  as shown in Fig. \ref{4AEI2m3m1V1gup1}(b) which is similar to Fig. 4(a)  of Ref. \cite{Ciccotti04} (for the AE amplitudes). [Fig. \ref{4AEI2m3m1V1gup1}(b)  corresponds to  Fig. \ref{I2m3V1gu1p1}(b).] In contrast, for  high pull velocities and low mass, and for a range of $\gamma_u$, ${R}_{ae}^s$ exhibits continuous bursts as shown in Fig. \ref{4AEI2m3m1V1gup1}(c),  which is again seen in experiments (Fig. 4(b)  of Ref. (\cite{Ciccotti04}). High mass and low pull velocity also exhibits continuous bursts. 

Denoting  $E_R$ to be the amplitude of ${R}_{ae}^s$ (i.e., from a maximum to the next minimum), for high pull velocities and low tape mass ( $C_f = 0.0078,v_c=0.0024,V^s = 5.93$), we find that the distribution of the magnitudes $D(E_R)$ shows a power law for all values of $\gamma_u$ investigated, i.e.,   $D(E_R)\sim E_R^{-m_E}$.   Further, $D(E_R)$ shown in Fig.\ref{4AEI2m3m1V1gup1}(d) exhibits two distinct scaling regimes as for case of  the distribution of the AE amplitudes ($A$) in experiments. The  value of $m_E \sim 0.6$ for the small amplitude regime while that for  large amplitudes that has a substantial scatter is about 1.8.  The  corresponding exponent values are  $m_A \sim 0.3$ and $3.2$ \cite{Ciccoae}. Using the fact that energy  $ E \propto A^2$, we get  $m_E = (1+m_A)/2$. Inserting  the values of $m_A$, we get the corresponding exponents  to be   $m_E \sim 0.65$ and $ m_E \sim 2.1$ which are close to the values  predicted by the model considering the scatter for the latter. In contrast, for high mass and low velocity case, we find a single scaling regime with an exponent $m_E \sim 0.69$.

Thus,  several qualitative features of the peel front dynamics observed in experiments are reproduced by the model. For example, the characteristic features of the AE signals observed in experiments, namely, noisy AE bursts   for low pull velocity changing over to continuous bursts at high pull velocity is  reproduced. For high pull velocities ( low tape mass), $D(E_R)$ exhibits two scaling regimes. However, comparison with experiments is made difficult due to the paucity of quantitative results except for the values of the exponents in the two scaling regimes which is in reasonable agreement with the model.  Even so, our study suggest  that if one wants a smooth peeling, one should peel at low velocity using high viscous dissipation. Significantly, the power law is seen at high pull speeds and thus unlike self organized criticality.  

The power law statistics for high pull velocities arises as a competition among the time scales due to inertia of the tape, dissipation  and imposed velocity  which is small at high $V^s$ leaving very little time for internal degrees of freedom to relax. The nature of peel front  ranges from synchronous peeling for large  $\gamma_u$ to rugged type for small $\gamma_u$.  The `stuck-peeled' configuration is  qualitatively similar to the inhomogeneous  peel fronts observed in experiments \cite{GL} as  also to the thin viscous film interface \cite{Shenoy}.    Interestingly, the 'verge of peeling picture' of the  peel front (Fig. \ref{I2m1V1gup01} b,c) is similar to the edge of unpinning picture of dislocations in the Portevin Le Chatelier effect \cite{Anan04}.  This is one of the few cases where the the power law emerges purely from deterministic dynamics (see \cite{Anan04}).

Here it is worth commenting on the assumption  that the integrand of Eq. 1 vanish  at each point $y$ which is valid when  $L >> b$ and when shear modulus $k_G$ is small. In principle, one should have a long range term of the form $k_G\int_0^b [u(y) - u(y')]^2dydy'/2\vert y - y'\vert $. An  equilibrium calculation with $u(y)$ defined at one end  shows that the shear strain energy is less than one percent compared to the total even for small $b/L =0.2$. This lends support for the PE term used. We have also carried out numerical calculations by retaining this term. For small $k_G$, results are not affected as it should be expected. However, for relatively high  values of $k_G$, the solutions that were smooth  breakup into stuck-peeled configurations. 

Finally, we note that properties of adhesive glue has been included in the model in an indirect way through the peel force function ( and low effective spring  constant of the peel front due to adhesive glue) and that of AE through the Rayleigh dissipation function. We state that while the model recovers most dynamical features of peeling,  issues that depend critically on the finite thickness of the adhesive  material (for instance, fibril formation)  cannot be addressed within the scope of the model.  

We thank Prof. A. J. Beaudoin for helpful discussions. G. A acknowledges the award of Raja Ramanna Fellowship and BRNS grant No. 2005/37/16/BRNS. 

\end{document}